\documentstyle[epsf,rotate,12pt]{article}

\newcommand{\beq}{\begin{equation}}
\newcommand{\eeq}{\end{equation}}

\begin{document}

\begin{center}
\vskip 2cm \Large {\bf Strong Phases in the Decays $B$ to
$\pi$$\pi$} \vskip 1.5cm \large Lincoln Wolfenstein and  Feng Wu
\vskip 1cm \small
{\em Department of Physics,\\
Carnegie-Mellon University\\
Pittsburgh, PA}

\end{center}
\vskip 2cm
\begin{abstract}

Two sources of strong phases in the decays $B$ to $\pi\pi$ are
identified: (1) ``quasi-elastic scattering" corresponding to
intermediate states like $\pi\pi$ and $\rho\rho$, (2)
``$c\bar{c}$'' corresponding to intermediate states like
$D\bar{D}$ and $D^{*}\bar{D}^{*}$. Possibilities of using data to
identify these two sources are discussed and illustrated. Present
data suggests both sources may be significant.
\end{abstract}
\thispagestyle{empty}

\newpage
\pagenumbering{arabic}

\section*{}

The decay of $B \to \pi\pi$ may be considered as due to the effective interaction
\begin{eqnarray}
\lambda_{u} d(u\bar{u} - t\bar{t})b + \lambda_{c} d (c\bar{c} - t \bar{t})b,
\end{eqnarray}
where
\begin{eqnarray*}
& &\lambda_u = V_{ub}^{*} V_{ud} =e^{i\gamma} A \lambda^3 {\mathrm sin\beta \over \mathrm sin\alpha}\\
& &\lambda_c= V_{cb}^{*}V_{cd}=- A\lambda^3.
\end{eqnarray*}
The decay amplitudes may be written, neglecting the small
electroweak penguins and assuming isospin invariance,
\begin{eqnarray}
&&-A(\pi^{+}\pi^{-})=Te^{i \delta_{T}} e^{i\gamma} + P e^{i \delta_{P}},\\
&&-\sqrt{2}A(\pi^{0}\pi^{0}) = Ce^{i\delta_{C}}e^{i\gamma}-Pe^{i\delta_{P}},\\
&&-\sqrt{2}A(\pi^{+}\pi^{0})=(Te^{i\delta_{T}}+Ce^{i\delta_{C}})e^{i\gamma},
\end{eqnarray}
where $\delta_{T}$, $\delta_{C}$, and $\delta_{P}$ are phases due
to the strong final state interaction. In terms of the isospin
analysis of the $\lambda_u$ terms, $T$ and $C$ may be replaced by
$A_2$ and $A_0$ and $(\delta_C,\delta_T)$ by $(\delta_2,\delta_0)$
\begin{eqnarray}
&&Te^{i\delta_{T}}=e^{i\delta_{0}}(A_{2} e^{i(\delta_2-\delta_0)}+A_0),\\
&&Ce^{i\delta_{C}}=e^{i\delta_{0}}(2A_{2}
e^{i(\delta_2-\delta_0)}-A_0).
\end{eqnarray}
The $T$ term is often referred to as the tree amplitude
corresponding to the $b$ quark decay into $u \bar{u} d$, while $P$
is called the penguin corresponding to a loop diagram dominated by
the virtual $t$ quark. However, as can be seen from Eq (1) there
is also a $t \bar{t}$ loop contributing presumably in a small way
to $T$. This is because we are using what is called the $c$
convention in contrast to the $t$ convention that separates off
the $t$ loop\cite{gronau}\cite{botella}. Note that $P$ makes a
contribution to the $I=0$ final state so that the complete
amplitude for $I=0$ (sometimes given the notation $A_0$) is here a
sum of the $A_0$ and $P$ terms.

If the final state scattering were purely elastic the phases
$\delta_2$ and $\delta_0$ would be $\pi\pi$ scattering phases in
accordance with the Watson theorem. In fact the final state
scattering is expected to be very inelastic and described by the
$N\times N$ $S$ matrix at the center-of-mass energy equal to the
$B$ mass and $J=0$. The simple Watson theorem can be applied only
to final states that are eigenstates of the $S$ matrix. In general
for any weak interaction operator $O$ the phase due to the strong
interaction for the amplitude $A_f$ is determined from
\begin{eqnarray}
{\mathrm Im} A_f= {\mathrm Im} \sum_i \langle f \vert S^{1/2}
\vert i \rangle \langle i \vert O \vert B \rangle. \label{af}
\end{eqnarray}

In the case of $A_0$ and $A_2$ or, equivalently $C$ and $T$,
determining the strong phases $(\delta_2, \delta_0)$ or
$(\delta_C, \delta_T)$ the intermediate states $i$ are $u\bar{u}
d(\bar{q})$ state where $\bar{q}$ is the spectator. These include
$\pi\pi$, $\rho\rho$, and many others. We refer to these as
``quasi-elastic" since the intermediate states include $\pi\pi$
and arise from the same quark set.

In the case of $\delta_P$ there are two classes of intermediate
states: once again there is a phase due to ``quasi-elastic"
rescattering which would yield a phase $\delta_{0P}$, but also
intermediate states of the form $c \bar{c} d (\bar{q})$ such as
$DD$ or $D^{*} D^{*}$ which would yield a phase
$\delta_{D}$\cite{these}. Thus the imaginary part of $P$,
${\mathrm Im} P$, can be written
\begin{eqnarray}
{\mathrm Im} P = {\mathrm Im} (\mathrm{quasi\textendash elastic})
+ {\mathrm Im}(c\bar{c}). \label{p}
\end{eqnarray}
In the limit that ${\mathrm Im}(c\bar{c})$ vanishes we have
\begin{eqnarray}
{\mathrm sin}\delta_P={\mathrm sin} \delta_{0P} \equiv {{\mathrm
Im} (\mathrm{quasi\textendash elastic}) \over |P|}. \label{0p}
\end{eqnarray}
In general
$\delta_{0P}$ and $\delta_0$ need not be equal although they
involve the same final $S$ matrix because the contributions of
different states $i$ to the sum in Eq (\ref{af}) may not be the
same for penguin and tree operators. There is also a contribution
of the form $s \bar{s} d (\bar{d})$ corresponding to states such
as $ K\bar{K}$; we expect this to be very small but include it in
$\delta_{0P}$.

We turn to the question of theoretical expectations for these two
types of strong phases. It is often stated that the outgoing
$\pi\pi$ pair do not scatter thus ruling out the quasi-elastic
source\cite{bjorken}. This is clearly wrong since reasonable
estimates give a significant value for the $\pi\pi$ cross-section
at the energy $5.3$ GeV. It can be argued that the multi-particle
states which dominate the $u\bar{u} d (\bar{q})$ final states in
$B$ decay are not likely to rescatter into the $\pi\pi$ state.
However since the final $\pi\pi$ states are less than one in a
thousand of the final states even a small rescattering can yield a
significant phase. It can be argued that different terms in the
sum in Eq (\ref{af}) cancel, but statistical analysis\cite{suzuki}
allows for a significant final phase in spite of this.

At first one might expect that the $c\bar{c}$ states would be
unimportant since scattering from these states to $\pi\pi$ is
expected to be small by the Zweig rule. Furthermore the total
branching ratio into these states is expected to be no more than a
factor of 2 larger than the total rate for the $u\bar{u}$ states
so the quasi-elastic would be much larger than the $c\bar{c}$
contribution. However two-body states are much more common among
the $c\bar{c}$ states and these are expected to rescatter more
readily into $\pi\pi$ than the multi-particle states that dominate
$u\bar{u}$. For example, the branching ratio to $D^{*}D^{*}$ is 30
to 40 times larger than that to $\rho \rho$. Thus a number of
papers have suggested that this should dominate the strong phase
\cite{barshay}.

We now turn to what we can learn from experimental results. We
assume the standard model and that the phase $\gamma$ has been
determined from other experiments. The experimental results are
the branching fraction ratios for the three $\pi\pi$ decays and
the asymmetries $C_{+-}$ and $S_{+-}$; recent experimental results
are summarized in Table 1\cite{heavy}. The values of $C_{+-}$ and
$S_{+-}$ can be used to determine $P/T$ and $\delta_{PT}\equiv
\delta_{P}-\delta_{T}$. In the approximation that $P/T$ is small
and $\beta+\gamma \textless 90^{\circ}$
\begin{eqnarray*}
\mathrm{tan}(\delta_{PT})\simeq
\mathrm{cos}(2(\beta+\gamma)){C_{+-} \over S^{'}},
\end{eqnarray*}
where $S^{'}=-(S+ \mathrm{sin}(2(\beta+\gamma)))$. Exact results
are shown in Table 2 for the central values in Table 1 and three
values of $\gamma$. As $\gamma$ becomes smaller $\delta_{PT}$
passes through $-90$ degrees. Phases with magnitudes greater than
$90^{\circ}$ would be interpreted as a final state strong phase
less than $90$ degrees with a reversal of the sign of $P/T$.

\begin{table}[tb]
\caption{} \label{tab:exp-data}
\begin{center}
\begin{tabular}{|l|c|c|c|}
\hline & $Br[10^{-6}]$& $C_{\pi\pi}$&$S_{\pi\pi}$
\\ \hline\hline
$B^{0}\to \pi^+\pi^-$& $4.8\pm 0.5$&$-0.37\pm 0.10$&$-0.50\pm0.12$
\\\hline
$B^{+}\to \pi^+\pi^{0}$& $5.6_{-1.1}^{+0.9}$& &
\\ \hline
$B^{0}\to \pi^{0}\pi^{0}$& $1.51\pm0.28$ &  &

\\ \hline
\end{tabular}
\end{center}
\end{table}

\begin{table}[tb]
\caption{Results for the central values in Table 1.}
\label{tab:exp-data}
\begin{center}
\begin{tabular}{|l|c|c|c|}
\hline $\gamma$ & $47^{\circ}$& $57^{\circ}$&$67^{\circ}$
\\ \hline\hline
${p \over T}$& 0.26&0.28&0.43
\\\hline
$\delta_{P}-\delta_{T}$& $-117.37^{\circ}$&$-68.97^{\circ}$ &
$-42.23^{\circ}$
\\ \hline
${C\over T}$& 0.587&0.768&0.906
\\
 & [0.883]&[0.970]&[1.027]
 \\ \hline
 $\delta_{C}-\delta_{T}$ & $+45.29^{\circ}$& $+34.46^{\circ}$ &
$+26.67^{\circ}$
\\
 &  $[-77.16^{\circ}]$&$[-61.83^{\circ}]$ &
$[-47.58^{\circ}]$
\\ \hline
${A_2\over A_0}$& 0.898&1.178&1.474
\\
 & [0.737]&[0.958]&[1.227]
 \\ \hline
$\delta_{2}-\delta_{0}$& $+31.18^{\circ}$&$+32.54^{\circ}$ &
$+31.53^{\circ}$
\\
 &  $[-61.26^{\circ}]$&$[-59.40^{\circ}]$ &
$[-54.24^{\circ}]$
\\ \hline
$\delta_{P}-\delta_{0}$& $-102.64^{\circ}$&$-51.33^{\circ}$ &
$-23.37^{\circ}$
\\
 & $[-142.88^{\circ}]$&$[-97.98^{\circ}]$ &
$[-72.34^{\circ}]$

\\ \hline
${C_{00}}$& 0.108&0.488&0.746
\\
 & [0.353]&[0.079]&[-0.085]

\\ \hline
\end{tabular}
\end{center}
\end{table}

The three $\pi \pi$ rates can now be used to determine $C/T$ and
$\delta_{CT} \equiv \delta_{C}-\delta_{T}$, or, equivalently,
$A_{2}/A_{0}$ and $\delta_{20} \equiv \delta_{2}-\delta_{0}$. This
is illustrated in Table 2 for the central values in Table 1. Using
these values and Eq (5) we determine $\delta_{T}-\delta_{0}$ and
then from $\delta_{PT}$ we obtain $\delta_{P0} \equiv
\delta_{P}-\delta_{0}$. Two solutions for $\delta_{CT}$ with
opposite signs are shown. The positive sign leads to lower and
more reasonable value for $\delta_{P0}$; also if the difference
between $\delta_{2}$ and $\delta_{0}$ is due to isovector exchange
in the rescattering one obtains the positive sign. The two
solutions may be distinguished by the values of $C_{00}$; limited
data available so far favors the negative sign for the central
value of $\gamma$.

The values for $\delta_{20}$ are entirely due to ``quasi-elastic"
rescattering. The value is seen to depend significantly on the
$\pi^{0}\pi^{0}$ rate as shown in Fig 1 and is quite large for the
present central value. If we assume that $\delta_{0P}\sim
\delta_{0}$ then in the absence of a $c\bar{c}$ contribution
$\delta_P \sim \delta_0$ (Eq (\ref{0p})); in this case the large
values for $(\delta_{P}-\delta_{0})$ shown in Table 2 would be
evidence for $c\bar{c}$ states contributing to $\delta_{P}$.
However it is important to note that the experiments cannot
determine $\delta_0$ or $\delta_{0P}$ or $\delta_0-\delta_{0P}$;
the isospin-independent quasi-elastic strong phase in $T$ cannot
be determined.

In conclusion we have tried to show what can be learned as to the
origin of the strong phases simply using data on $B$ to $\pi\pi$
decays. A more ambitious attempt including $B$ to $K\pi$ and using
$SU(3)$ has been made by Christopher Smith\cite{smith}. Similar results
to those shown in Table 2 have been given in a number of
papers\cite{ya} devoted to determining $\gamma$. Here we do not try
to analyze the data in detail but simply try to illustrate what
strong phase information can be obtained. Our conclusions are:

(1) Definite information on quasi-elastic strong phases in the
tree amplitude can be obtained and present data points to a value
$\delta_{20}$ of order $25^{\circ}$ or larger but this is very
sensitive to the $\pi^{0}\pi^{0}$ branching ratio.

(2) A second strong phase $(\delta_P-\delta_0)$, which appears to
be quite large, is associated with the penguin and has in general
both quasi-elastic and $c\bar{c}$ contributions. If
$(\delta_{0P}-\delta_{0})$, which represents the difference
between the quasi-elastic $I=0$ phase for the penguin and that for
the tree, is small then there must be a significant $c\bar{c}$
term. This represents the contribution of rescattering  from
$c\bar{c}$ states like $D\bar{D}$ or $D^{*}\bar{D}^{*}$. Thus
present data suggests that both sources of strong phases may be
significant.

\section*{Acknowledgements}
One of us (LW) wants to thank Jo\~{a}o Silva for helpful
discussions and for the hospitality of the Instituto Superior
T\'{e}cnico in Lisbon where this work was started. This research
was supported in part by the Department of Energy under no.
DE-FG02-91ER40682.

\newpage
\begin{figure}[t]
\begin{minipage}{7.0cm}
\centerline{\epsfxsize=10.0cm  \epsffile{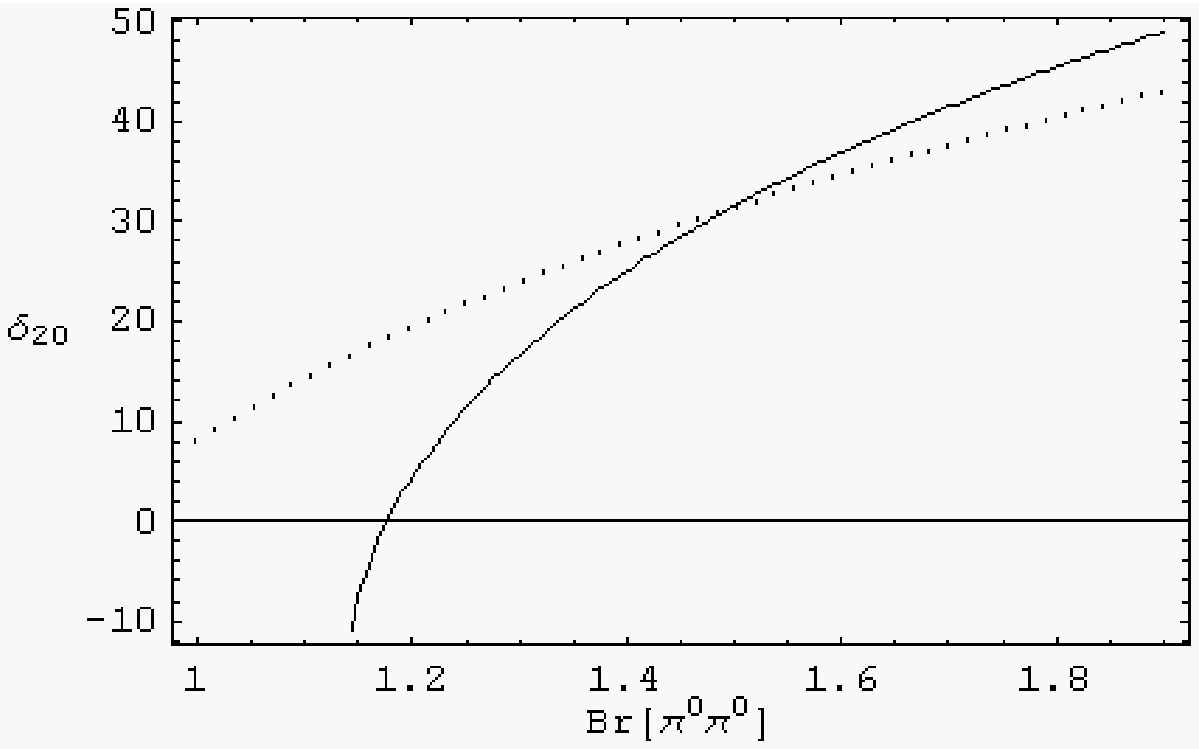} } \vspace{0cm}
\end{minipage}
\hfill
\begin{minipage}{7.0cm}
\centerline{\epsfxsize=9.5cm  \epsffile{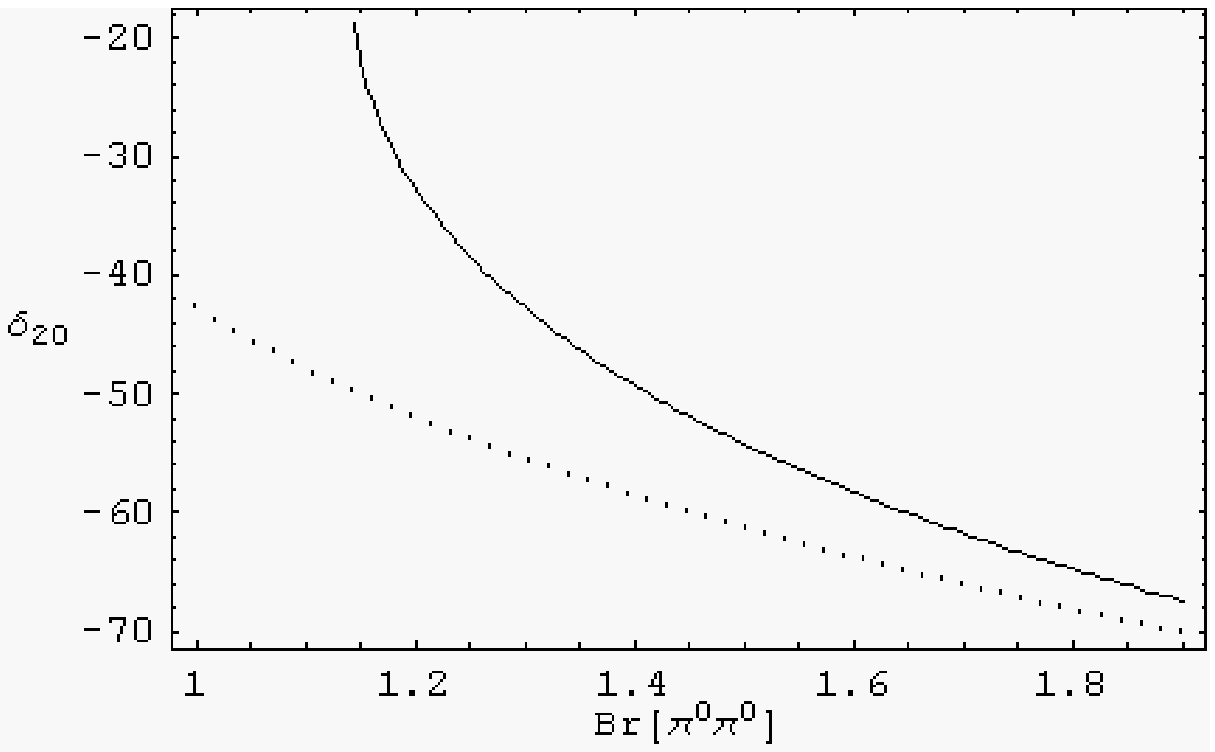} } \vspace{0cm}
\end{minipage}
\caption{The two solutions for the strong phase difference
$\delta_{20}$ as a function of the $\pi^{0}\pi^{0}$ branching
ratio using central values for the other observables and two
values of $\gamma$ (Solid line $\gamma=67^{\circ}$, Dashed line
$\gamma=47^{\circ}$).}
\end{figure}

\end{document}